\documentstyle[12pt]{article}
\newcommand{\BB}{{\cal B}}
\newcommand{\MM}{{\cal M}}

\newcommand{\vu}{\vec u}

\newcommand{\be}{\begin{equation}}
\newcommand{\ee}{\end{equation}}
\newcommand{\ben}{\begin{eqnarray}\displaystyle}
\newcommand{\een}{\end{eqnarray}}
\newcommand{\refb}[1]{(\ref{#1})}

\begin{document}

{}~ \hfill\vbox{\hbox{hep-th/9702165}\hbox{MRI-PHY/P970202}
\hbox{NI 97013}}\break

\vskip 3.5cm

\centerline{\large \bf Orientifold Limit of F-theory Vacua}

\vspace*{6.0ex}

\centerline{\large \rm Ashoke Sen\footnote{On leave of absence from 
Tata Institute of Fundamental Research, Homi Bhabha Road, 
Bombay 400005, INDIA}
\footnote{E-mail: sen@mri.ernet.in, sen@theory.tifr.res.in}}

\vspace*{1.5ex}

\centerline{\large \it Mehta Research Institute of Mathematics}
 \centerline{\large \it and Mathematical Physics}

\centerline{\large \it  Chhatnag Road, Jhoosi,
Allahabad 221506, INDIA}
\vspace*{1.5ex}

\centerline{\large \it and}
\vspace*{1.5ex}

\centerline{\large \it Issac Newton Institute for Mathematical Sciences}
\centerline{\large \it University of Cambridge}

\centerline{\large \it Cambridge, CB3 0EH, U.K.} 

\vspace*{4.5ex}

\centerline {\bf Abstract}

We show how an F-theory compactified on a Calabi-Yau $(n+1)$-fold
in appropriate weak coupling
limit reduces formally to an orientifold of type IIB theory
compactified on an auxiliary complex $n$-fold. In some cases
(but not always)  
if the original $(n+1)$-fold is singular, then the auxiliary $n$-fold is
also singular.
We illustrate this by analysing 
F-theory on elliptically fibered Calabi-Yau 3-folds on base $F_n$.

\vfill \eject

\baselineskip=18pt


F-theory\cite{VAFAF,FTHEORY}
and orientifolds\cite{ORIENT,DBRANE,BIAN,GIMPOL}
are two different classes of compactifications of type
IIB string theory. In recent years we have seen several examples 
of dual pairs of theories, of which one is an F-theory compactification,
and the other is an 
orientifold\cite{SENF,BLUZAF,DABPAR,GIMJOH,GOPMUK,PARK,SGIM}.
Typically in these examples
the F-theory background describes the non-perturbative 
correction to the orientifold background\cite{SENF,BDS,PREV,SGIM}.

In this note we shall show that this feature is quite general, $-$ namely
given any F-theory compactification on a Calabi-Yau $(n+1)$-fold, 
we can go to appropriate region in
the moduli space of the theory where the background (at least formally)
looks like that of
an orientifold of type IIB theory on an $n$-dimensional complex
manifold.  We start from the Weierstrass form of
an elliptically fibered manifold:
\be \label{e1}
y^2 = x^3 + f(\vu) x + g(\vu)\, ,
\ee
where $f$ and $g$ are appropriate polynomials on some base $\BB$
labelled by complex coordinates $\vu$. F-theory compactified on
such a manifold is by definition type IIB theory compactified on
the base $\BB$ with background axion-dilaton
field $\lambda(\vu)$ given by:
\be \label{e4}
j(\lambda) = {4 \cdot (24 f)^3\over 4f^3 + 27 g^2}\, ,
\ee
where $j(\lambda)$ is the modular invariant function of $\lambda$
with a pole at $i\infty$, zero at $e^{i\pi/3}$, and normalized such
that $j(i)=(24)^3$. The locations of
the seven branes on the base are at the zeroes of
\be \label{e5}
\Delta = 4 f^3 + 27 g^2\,  
\ee
where $\lambda\to i\infty$ up to an SL(2,Z) transformation.
We shall first consider a special family of points in the moduli space of
this F-theory where $f$ and $g$ take special form\cite{SGIM}:
\be \label{e6}
f(\vu) = C\eta(\vu) - 3 h(\vu)^2\, ,
\ee
and 
\be \label{e7}
g(\vu)= h(\vu) (C \eta(\vu) - 2 h(\vu)^2)\, .
\ee
Here $h$ and $\eta$ are appropriate polynomials on $\BB$, and $C$ is a
constant denoting the overall normalization of $\eta$. This gives
\be \label{e9}
j(\lambda) = {4\cdot (24)^3 (C\eta -3h^2)^3\over C^2 \eta^2 (4C\eta -9h^2)}
\, ,
\ee
and 
\be \label{e8}
\Delta = C^2 \eta^2 (4C\eta - 9h^2)\, .
\ee
{}From eq.\refb{e9} we see that as $C\to 0$ with $\eta$ and $h$
fixed, $j(\lambda)$ goes to 
infinity almost everywhere on $\BB$ except at the zeroes of the
numerator. This corresponds to $\lambda\to i\infty$ (up to an SL(2,Z)
transformation) almost everywhere
on $\BB$, and choosing the convention
that this limit is $i\infty$ and not one of its images under SL(2,Z),
we can identify this as the weak coupling limit.
In particular the average $\lambda$ is related to $C$ as
\be \label{ejj1}
e^{2\pi i\langle\lambda\rangle} \sim C^2\, ,
\ee
for small $C$.
We shall now show that in this limit the background $\lambda$ given
in \refb{e9} can be identified to that of an orientifold, with orientifold
seven plane situated at $h=0$, and a pair
of Dirichlet seven branes situated at $\eta=0$. 

To see this first of all
note from \refb{e8} that at $\eta=0$, $\Delta$ has a double zero.
Since neither $f$ nor $g$ vanish there, this is an $A_1$ 
singularity,\footnote{We are implicitly assuming that $\eta$ and $h$ do
not have any common factor, and that each has only simple zeroes.
Otherwise we might get more complicated 
singularities. We shall encounter such examples later.} and $\lambda\to
i\infty$ on this hypersurface up to an SL(2,Z) transformation. However, our
previous convention that for small $C$, $Im(\lambda)$ is large at a
generic point in $\BB$, and the fact that we can pass from a generic point
in $\BB$ to the $\eta=0$ surface keeping $j(\lambda)$ given in
\refb{e9} always large, shows that on the surface $\eta=0$
$\lambda$ actually goes to $i\infty$ and not
to any of its SL(2,Z) transform. 
The SL(2,Z) monodromy around the hypersurface $\eta=0$ is then 
given by $T^2$ where
\be \label{ee3}
T = \pmatrix{1 & 1\cr 0 & 1}\, .
\ee
Thus the singularity at $\eta=0$ can be
interpreted as due to the presence of a pair of coincident D-7 branes.

The other zeroes of $\Delta$ are situated at 
\be \label{e11}
9 h(\vu)^2 - 4 C\eta(\vu)=0\, ,
\ee
which can be rewritten as
\be \label{e13}
h(\vu) = \pm {2\over 3}\sqrt{C\eta(\vu)}\, .
\ee
For small $C$, this represents a pair of hypersurfaces close to the
hypersurface $h=0$.
The monodromy around each of these hypersurfaces is conjugate to $T$.
Let us take them to be\footnote{Instead of characterizing these
hypersurfaces by the SL(2,Z) monodromy around them, we could also
characterize them by the value of $\lambda$ on the hypersurface. 
A monodromy matrix $MTM^{-1}$ with $M=\pmatrix{p & q\cr r & s}$
will correspond to $\lambda$ being equal to $p/r$ on that hypersurface.}
\be \label{e14}
MTM^{-1} \quad \hbox{and} \quad NTN^{-1}\, ,
\ee
respectively, where $M$ and $N$
are some SL(2,Z) matrices. 
Thus the effect of going around both these branches is given by
$MTM^{-1}NTN^{-1}$. We shall now explicitly compute this monodromy from
\refb{e9}. As seen from this equation, for small $C$, 
$j(\lambda)$ is large everywhere along a contour 
enclosing the hypersurface
$h=0$ {\it as long as the contour remains at a finite distance from this
hypersurface, thereby enclosing both branches of
\refb{e13}}. Thus $Im(\lambda)$ remains large along this contour and comes
back to its original value as we travel once around the contour. 
Since $j(\lambda)\sim exp(-2\pi i\lambda)$ for large $Im(\lambda)$,
the change in $Re(\lambda)$ is given by
\be \label{enew1}
-{1\over 2\pi i} \ointop {d j\over j}\, ,
\ee
where the integral is performed along
the contour around $h=0$. For small $C$, \refb{e9} gives
\be \label{enew2}
j(\lambda)\sim h^4/C^2 \eta^2\, .
\ee
Substituting this in \refb{enew1} and picking up the contribution from
the pole at $h=0$, we see that $Re(\lambda)$ changes by $-4$ along this
contour.\footnote{Since we are interested in calculating the monodromy
around the two branches of the hypersurface given in \refb{e13} we choose
the contour in such a way that it does not enclose any branch of the
hypersurface $\eta=0$.} Thus we get
\be \label{e17}
MTM^{-1} NTN^{-1} = \pm T^{-4}\, .
\ee
Starting with the most general ansatz for the SL(2,Z) matrices $M$ and $N$,
and substituting in eq.\refb{e17}, one can verify that the most general
solutions for $MTM^{-1}$ and $NTN^{-1}$ are of the form:
\be \label{e18}
MTM^{-1} = \pmatrix{1-p & p^2 \cr -1 & 1+p}\, ,
\qquad NTN^{-1} = \pmatrix{-1-p & (p+2)^2 \cr -1 & 3+p}\, ,
\ee
where $p$ is an arbitrary integer.
This gives: 
\be \label{e19}
MTM^{-1} NTN^{-1} = - T^{-4}\, .
\ee
Now since in the weak coupling limit $C\to 0$ the two branches collapse
onto the surface $h=0$, only this monodromy will be visible in this
limit. But this is precisely the monodromy around an orientifold
seven plane. In order to describe this in more concrete terms,
let us consider an auxiliary manifold $\MM$
which is a double cover of the 
base $\BB$, and is defined by the equation:
\be \label{e20}
\xi^2 - h(\vu) = 0\, ,
\ee
where $\xi$ is a complex variable. We now consider type IIB theory
on this auxiliary manifold $\MM$, and mod out this theory
by the transformation
\be \label{e21}
(-1)^{F_L}\cdot\Omega\cdot\sigma \, ,
\ee
where $(-1)^{F_L}$ is the discrete
$Z_2$ symmetry of type IIB theory that changes the sign of all the
Ramond sector states in the left-moving sector of the
world-sheet, $\Omega$ is the world-sheet parity
transformation, and $\sigma$ denotes the transformation 
\be \label{e22}
\xi \to -\xi\, .
\ee
This transformation leaves fixed the seven plane $\xi=0$, which using
\refb{e20} can also be written as $h(\vu)=0$. This represents the
orientifold plane.
Then a closed curve around this  plane in the 
quotient manifold $\BB$ 
will have precisely the monodromy given in \refb{e19}. 
The $-1$ factor in the monodromy
\refb{e19} is the effect of the transformation $(-1)^{F_L}\cdot\Omega$,
whereas the $T^{-4}$ factor reflects the fact that such an orientifold
seven plane carries $-4$ units of magnetic charge of the Ramond-Ramond
scalar field that forms the real part of $\lambda$\cite{GIMPOL,SENF}.
The splitting of the orientifold plane into two seven-branes for non-zero
$C$ reflects the phenomenon already observed in ref.\cite{SENF}.

This establishes that the F-theory background, in the particular limit
we have described, does represent an orientifold of IIB on $\MM$ with
a pair of D-branes situated at $\eta(\vu)=0$ on the quotient $\BB$.
We shall now consider deformation of \refb{e6}, \refb{e7} to most
general $f$ and $g$. Since $\eta$ is an arbitrary polynomial (subject
to the same restrictions as $f$, and its overall normalization specified
separately in the constant $C$), we see from \refb{e6} that by 
choosing the most general $\eta$ and $C$ we can get the most general $f$.
Thus we only need to deform $g$. To this end consider the following form
of $g$:
\be \label{epp1}
g = h(C\eta - 2h^2) + C^2 \chi\, ,
\ee
where $\chi$ is a general polynomial subject to the same restrictions
as $g$. This will certainly give us the most general $g$.
Of course some of the deformations of $\chi$ will simply correspond
to deformations of $h$ and $\eta$ keeping $f$ fixed, but this will not
concern us here as we are only interested in showing that we have the
most general $g$. The factor of $C$ in \refb{epp1} has been chosen
such that $\Delta$ still has an overall factor of $C^2$ and the relation
\refb{ejj1} remains valid. Indeed with this choice,
\be \label{epp2}
\Delta = C^2 \{\eta^2 (4C\eta - 9 h^2) + 54 h (C\eta-2h^2)\chi
+ 27 C^2 \chi^2\}\, .
\ee
This no longer has a factorized form.
In order to identify the orientifold planes and D-branes, we need to take
the weak coupling limit $C\to 0$ keeping $h$, $\eta$ and $\chi$
fixed. In this limit:
\be \label{epp3}
\Delta = -9C^2 h^2 (\eta^2 + 12 h\chi)\, .
\ee 
{}From this we see that the orientifold plane is still located at
$h=0$, but the D-brane positions have shifted to
\be \label{epp4}
\eta = \pm \sqrt{-12 h \chi}\, .
\ee
Thus the D-branes are split, but they coincide at the point of
intersection with the orientifold plane. Also as we travel once around the
orientifold plane $h=0$ remaining on the  surface of the
D-brane, we see from \refb{epp4} that the two D-branes get exchanged.
This is entirely in agreement with the results found in \cite{PREV}.
The splitting of the D-brane pair can be described, in the orientifold
language, as a result of switching on vacuum expectation values of
fields that are charged under the SU(2) gauge group associated with the
original D-brane pair.

We would also like to know if the complex $n$-fold defined by 
eq.\refb{e20} satisfies the Calabi-Yau condition. 
Due to physics
reasons we know that this must be the case, since otherwise type IIB
compactified on \refb{e20} will not have any supersymmetry, and hence
its orientifold will also not have any unbroken space-time supersymmetry.
But we can also try to verify this directly. First let us
consider the case where the base is a toric variety. In this case we
can introduce homogeneous coordinates $(u_1,\cdots u_m)$ on the base
with identifications of the form:
\be \label{ez1}
(u_1, \cdots u_m) \equiv \big((\lambda^{(i)})^{w_1^{(i)}}u_1, \cdots 
(\lambda^{(i)})^{w_m^{(i)}}u_m\big) \quad \hbox{for}\quad 1\le i\le p\, ,
\ee
where $\lambda^{(i)}$ for $1\le i\le p$ are non-zero complex numbers.
This describes a toric variety of complex dimension $n=(m-p)$.  Let us use
the notation that the vector $\vec w_k\equiv (w_k^{(1)}, \ldots
w_k^{(p)})$ denotes the weight of the $k$th coordinate $u_k$.
$f$ and $g$ appearing in eq.\refb{e1} must be homogeneous polynomials
of the coordinates $u_k$ with weight $2\vec w$ and $3\vec w$ respectively
for some $p$ dimensional vector $\vec w$, so that we can assign weights
$\vec w$ and $(3\vec w/2)$ to the coordinates $x$ and $y$ respectively.
In order that \refb{e1} describes a Calabi-Yau $(n+1)$-fold, the total
weight of the coordinates $x$, $y$ and the $u_k$'s must be equal to the
weight of the polynomial on the left hand side of \refb{e1}. This gives
\be \label{ez2}
3\vec w = \vec w + {3\over 2} \vec w + \sum_{k=1}^m \vec w_k\, .
\ee
Let us now turn to eq.\refb{e20}. From eqs.\refb{e6} and \refb{epp1} we
see that $\chi$, $\eta$ and $h$ must be homogeneous polynomials in $\{u_k\}$
with weights $3\vec w$, $2\vec w$ and $\vec w$ respectively, and hence $\xi$ 
in eq.\refb{e20} must have weight $\vec w/2$. In order that \refb{e20}
describes a Calabi-Yau manifold, the sum of weights of $\xi$ and the
$u_k$'s must be equal to that of the polynomial on the right hand
side of \refb{e20}. This requires
\be \label{ez3}
\vec w = {1\over 2}\vec w + \sum_{k=1}^m \vec w_k\, .
\ee
But this is simply a consequence of \refb{ez2}. Thus we see that if the
original $(n+1)$-fold that we started with describes a 
Calabi-Yau $(n+1)$-fold,
then the auxiliary $n$-fold \refb{e20} also satisfies the Calabi-Yau
condition. Of course we shall also need to analyse the possible singularities
of this $n$-fold in each case separately.

We can also give a more general argument that does not depend on the base
being a toric variety.  For this let us
take $f$ and $g$ appearing in \refb{e1} to be sections of line bundles
$L^{\otimes 4}$ and $L^{\otimes 6}$ respectively for some line bundle
$L$ on $\BB$, so that $x$ and $y$ appearing in \refb{e1} can be regarded as
taking values in $L^{\otimes 2}$ and $L^{\otimes 3}$ respectively.
In order that \refb{e1} describes a Calabi-Yau manifold, its first
Chern class must vanish. This requires,
\be \label{ezy1}
c_1(\BB) + c_1(L)(3 + 2 -6)=0\, .
\ee
The coefficients 3, 2 and $6$ of $c_1(L)$ on the left hand side
of this equation
reflect the degree of $y$, $x$ and the constraint
\refb{e1} respectively. (Here by degree we simply refer to
the power of $L$ that appears in the corresponding line bundle). This
gives
\be \label{ezy2}
c_1(\BB) = c_1(L)\, .
\ee
Let us now turn to eq.\refb{e20}. From eqs.\refb{e6} and \refb{epp1} we
see that $\chi$, $\eta$ and $h$ must describe sections of the line bundles
$L^{\otimes 6}$, $L^{\otimes 4}$ and $L^{\otimes 2}$ respectively, and
hence $\xi$
in eq.\refb{e20} must be valued in the line bundle $L$.
The condition for the vanishing of the
first Chern class of the manifold described by \refb{e20} is
then given by,
\be \label{ezy3}
c_1(\BB) + c_1(L) (1 - 2) =0\, .
\ee
Again the coefficients 1 and 2 of $c_1(L)$ on the left hand side of
\refb{ez3} reflect the degree of $\xi$ and the constraint
\refb{e20} respectively. But \refb{ezy3} is simply a consequence
of \refb{ezy2}.  Thus the auxiliary $n$-fold \refb{e20} satisfies 
the Calabi-Yau condition. 

Let us now illustrate this in the context of a class of F-theory
compactifications discussed in ref.\cite{FTHEORY}, namely on elliptically
fibered Calabi-Yau 3-folds on base $F_n$. 
Let $(u,v)$ denote the affine coordinates on the base.
Then the polynomials
$f$ and $g$ appearing in eq.\refb{e1} are of the form\cite{FTHEORY}:
\be \label{e2}
f(u,v) = \sum_{k=0}^8 \sum_{l=0}^{8-4n+nk} f_{kl} u^k v^l\, ,
\ee
and,
\be \label{e3}
g(u,v) = \sum_{k=0}^{12} \sum_{l=0}^{12-6n+nk} g_{kl} u^k v^l\, .
\ee
Thus we can take
\be \label{e7b}
h(u,v) = \sum_{k=0}^4 \sum_{l=0}^{4-2n+nk} h_{kl} u^k v^l\, ,
\ee
\be \label{e7a}
\eta(u,v) = \sum_{k=0}^8 \sum_{l=0}^{8-4n+nk} \eta_{kl} u^k v^l\, ,
\ee
\be \label{exx1}
\chi(u,v) = \sum_{k=0}^{12} \sum_{l=0}^{12-6n+nk} \chi_{kl} u^k v^l\, ,
\ee
so that $f$ and $g$ given in eqs.\refb{e6} and \refb{epp1} are of the
form given in \refb{e2}, \refb{e3}. 
Let us now try to
determine under what condition the surface \refb{e20} with $h$
as given in \refb{e7b} is non-singular. 
For this we
note from \refb{e7b} that for a given $n$, $h_{kl}$ is non-zero only
for
\be \label{ekk1}
k\ge 2 - {4\over n}\, , \qquad l\le 4-2n+nk\, .
\ee
Thus for $n\le 2$, $h_{00}$ is non-vanishing, and hence the surface
defined in eq.\refb{e20} has no singularity at $u=v=0$. On the other
hand for $n\ge 5$, $h_{0l}$ and $h_{1l}$ vanish for all $l$, and 
\refb{e20} takes the form:
\be \label{ekk2}
\xi^2 - u^2\sum_{l}h_{2l} v^l + O(u^3) = 0\, \ee
This surface is clearly singular at $u=\xi=0$ for all $v$ since
the the left hand side of eq.\refb{ekk2}, as well as all its 
derivatives vanish at $u=\xi=0$. This is related to the fact that our
original assumption that $h$ has only simple zeroes 
breaks down in this case.

Thus it remains to analyse the two cases $n=3$ and $n=4$. For $n=4$
the surface takes the form:
\be \label{ekk3}
\xi^2 - h_{10} u + O(u^2)=0\, .
\ee
Thus the derivative of the left hand side of this equation with
respect to $u$ does not vanish at $u=\xi=0$ and the surface is non-singular
there. For $n=3$, on the other hand, the equation of the surface is
\be \label{ekk4}
\xi^2 - u (h_{10} + h_{11} v) + O(u^2)=0\, .
\ee
In this case the left hand side of this equation, as well as all its
derivatives vanish at $\xi=u=0$, $v=-h_{10}/h_{11}$, and the surface
is singular at that point.

Thus the manifold $\MM$ is non-singular for $n=0,1,2$ and 4. The $n=2$
case is already known to be equivalent to the $n=0$ case\cite{FTHEORY},
so we have three independent cases. These are precisely the cases for
which the Calabi-Yau manifold elliptically fibered over $F_n$ were
found to be equivalent to Voicin-Borcea orbifolds\cite{VOI,BOR} in 
ref.\cite{FTHEORY}. Our construction makes the orientifold limit of
these models explicit, and at the same time provides us with the general
configuration of D-branes allowed for this orientifold, 
namely on the 
hypersurface \refb{epp4} with $h$, $\eta$ and $\chi$ as given in
eqs.\refb{e7b}-\refb{exx1}.
Note also
that for $n=4$, $h\sim u$, $\eta\sim u^2$ and $\chi\sim u^3$
for small $u$. Thus there
is a $D_4$ singularity at $u=0$. From the orientifold viewpoint this
corresponds to four D-branes on top of an orientifold plane at $u=0$.
{}From the general form of $h$, $\eta$ and $\chi$
it can easily be seen that the
$u=0$ plane does not intersect any other component of D-branes or
orientifold planes; hence the unbroken gauge group at a generic
point in the moduli space is $SO(8)$\cite{FTHEORY}. 

In special cases one may be able to take the limit where
the auxiliary manifold $\MM$ itself can be regarded 
as an orbifold of a torus.
In this case the model is mapped to an orientifold of type IIB 
compactified on a torus, as was the case 
in refs.\cite{SENF,BLUZAF,DABPAR,SGIM}. However we should add a
cautionary remark here. Typically conformal field theory orbifolds 
do not correspond to geometric orbifolds, but represent geometric
orbifolds accompanied by half unit of background $B_{\mu\nu}$ 
flux through the collapsed two cycles\cite{ASPINK3}. On the other hand
in F-theory we do not normally have this flux. Thus these two
theories share the same background axion-dilaton field, but different
background tensor fields. If the tensor field flux is even under the
orientifold projection (as is the case in ref.\cite{SGIM}), then 
the two are in the same moduli space, and we can continuously go
from the orientifold to the F-theory by switching off the tensor field
flux. On the other hand if the tensor field flux is odd under the
orientifold projection (as is the case in the model of
refs.\cite{BLUZAF,DABPAR,COSH})
then we cannot switch it off continuously and the two theories are
in different moduli spaces in general. 
(Note that half unit of tensor field flux
is still even, and hence will survive the orientifold projection.)

The results of this paper can also be used in reverse, namely
to construct orientifolds
of type IIB theory compactified on a general Calabi-Yau $n$-fold with
a $Z_2$ isometry that leaves fixed a surface of codimension one. We can
mod out the theory by a combination of this $Z_2$ transformation and
$(-1)^{F_L}\cdot\Omega$ to get an orientifold. Let $\vu$ denote the 
complex coordinates on
the quotient, and $h(\vu)=0$ denote the location of the orientifold
plane in this quotient. In order to cancel the RR charge carried by 
the orientifold plane, we need to place appropriate D-branes on this
quotient manifold. This can be done by placing 
the Dirichlet 7-branes along the surface $\eta^2+12h\chi=0$, 
where $\eta$, $h$ and $\chi$
satisfy the condition that the surface described in \refb{e1} with
$f$ and $g$ given in eqs.\refb{e6}, \refb{epp1}, describes a Calabi-Yau
$(n+1)$-fold. $\lambda$ given in \refb{e4} provides a non-perturbative
description of the background around such a configuration of D-brane
pairs and orientifold planes.
For $n=3$, we also need to place appropriate number of
three branes filling non-compact part of space-time to cancel all the
tadpoles\cite{SETV}.

\noindent{\bf Acknowledgement}:  I wish to thank 
S. Yankielowicz for useful discussions. This work was supported in part by a
grant from NM Rothschild and Sons Ltd.

\end{document}